\documentclass{aa}
\usepackage{graphicx}
\begin{document}
   \title{A close look into an intermediate redshift galaxy using STIS}


   \author{Du\'{\i}lia F. de Mello
          \inst{1}
          \and
          Anna Pasquali
	  \inst{2}
          }

   \offprints{duilia@oso.chalmers.se or apasqual@eso.org}

   \institute{Onsala Space Observatory, Chalmers University of Technology,
              SE- 43992 Onsala, Sweden\\
              \email duilia@oso.chalmers.se
         \and
             ESO/ST-ECF,
	     Karl-Schwarzschild-Strasse 2, 85748 
	     Garching bei Muenchen,Germany\\
             \email apasqual@eso.org}

   \date{Received; accepted}

   \abstract{We present a detailed view of a galaxy at $z=0.4$ which is part of a 
   large database of intermediate redshifts using high resolution images. We used the 
   STIS parallel images and spectra to identify the object and obtain 
   the redshift. The high resolution STIS image (0.05$''$) enabled us to analyse the
   internal structures of this galaxy. A bar along the major axis and hot-spots of 
   star formation separated by 0.37$''$ (1.6 kpc) are found along the inner region 
   of the galaxy. The analysis of the morphology of faint galaxies like this one is an 
   important step towards estimating the epoch of formation of the Hubble classification 
   sequence.
   }

\maketitle
%

\section{Introduction}

The attempts to connect the luminosity density in high-redshift galaxy 
samples to lower-redshift surveys (e.g. Madau et al. 1996, Steidel et al. 1999) suggest 
that star formation density increases rapidly from $z=0$ to $z=1$. However, despite the efforts made so far, 
the nature of the objects and the mechanisms which are the drivers of galaxy 
evolution at intermediate redshifts are still under debate. The strong 
clustering of faint galaxies at intermediate redshifts (Infante et al. 1995, 
Carlberg et al. 2001) suggests that mergers were more frequent at that time 
than in the present and would be responsible for an increase of the star formation at those epochs.
It is also possible that there is an excess of dwarf galaxies having their 
first burst of star formation at intermediate redshifts (Babul \& Ferguson 1996, Guzm\'an
1998). Therefore, a census of the galaxy population as a function of redshift would be 
ideal in order to better understand the nature of galaxies at different epochs. 
The analysis of the morphology of faint galaxies is a key step towards estimating the epoch of 
formation of the Hubble classification sequence. Deep exposures such as the Hubble Deep Field 
(HDF) campaigns (Williams et al. 1996, Williams et al. 2000) have enabled morphology classification 
of faint galaxies for the first time (e.g. van den Bergh et al. 1996). 
However, larger samples are required in order to draw an unbiased view of galaxy evolution. For 
instance, in the sample of Abraham et al. (1999), which includes the HDF and the HDF-South there
are only 15 spiral systems at $z<0.5$. Our aim is to improve this situation by assembling a 
statistically large database of galaxies at intermediate redshifts using high resolution images.  
In this letter we present the first field we have analysed.

\section{The Data - Why STIS?}

We are selecting our sample from the HST Space Telescope Imaging 
Spectrograph (STIS) high latitude fields, which have been observed in parallel mode. 
STIS is used in parallel with any other HST instrument 
selected as prime instrument for a specific program (GO). Therefore, STIS points to a
field adjacent to the GO and it is programmed to acquire direct
images (in the CLEAR Filter covering from 2000 \AA\/ to 1$\mu$m) 
and slitless spectra (with the G750L grating) of the parallel field, with exposure times which fit the
GO integrations. Direct images are characterized by a spatial sampling
of 0.05$''$, {\it the highest} offered by HST, and cover a field of view of ~51$''$ 
$\times$ 51$''$. They are used to derive the morphology of the galaxies in the field  
and to analyse their environments. The STIS slitless spectra of the same
field (with a spectral resolution of 4.9 \AA/pix) 
over the spectral range 5230--10270 \AA\ are extracted and searched for either emission
lines or Balmer jump which will define the redshift of the galaxies in the
parallel field.

The data presented in this letter were acquired with the HST/STIS spectrograph in 
parallel mode (Programme ID = 8549, PI: Baum) for a field at 
RA (2000) = 13$^{\rm h}$ 37$^{\rm m}$ 56$^{\rm s}$.9 and DEC (2000) = 70$^{\rm o}$ 16$'$ 54$''$.8. 
Two pairs of direct and slitless-spectra images were taken in February 2000; 
the individual fits files are summarized in Table 1 together with their 
exposure times. The primary instrument was WFPC2 dedicated to the imaging 
of the cluster Abell 2218 (Programme ID = 8500, PI: Fruchter). Abell 2218
($z=0.17$) is at RA (2000) = 16$^{\rm h}$ 35$^{\rm m}$ and DEC (2000) = 66$^{\rm o}$ 12$'$ (Smail et al. 2001) and,
therefore, not included in the parallel images which are $\sim$ 3 hours to the west and 4 
degrees to the north of Abell 2218.

\begin{table}
\caption[]{Log of the observations}
\begin{tabular}{cc|cc}
\hline
Images   & Exp. Time & Spectra & Exp. Time \\
         & s & & s\\
\hline
o5lil0sgq & 642 & o5lil0siq & 900\\
o5lil0slq & 150 & o5lil0smq & 1397\\
\hline
\end{tabular}
\end{table}

\begin{figure}
\centering
\includegraphics[width=8cm]{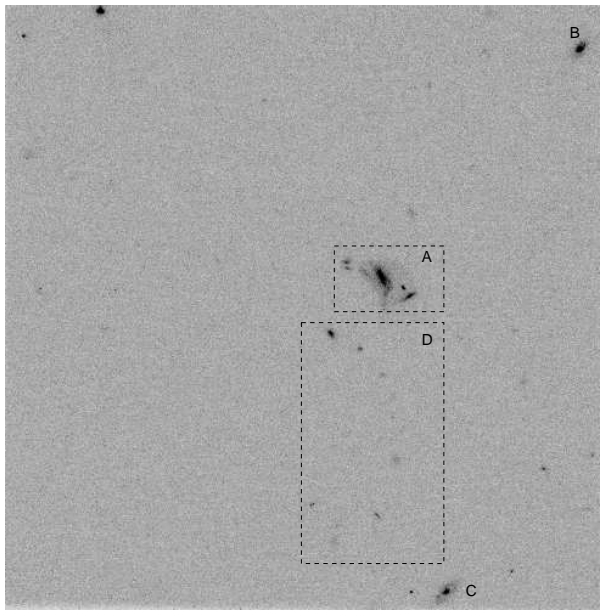}
\caption{STIS imaging of the field 204+70.2. Size of the field is 51$'' \times 51''$.
Dashed box A is shown in detail in Fig.~\ref{f4}.
North is to the top and East to left.}
\label{f1}
\end{figure}

The direct images presented here were acquired in the MIRVIS configuration, with the 
50CCD aperture and the CLEAR filter (Fig.~\ref{f1}). A number of pointings adjacent 
to the field are available in the archive, but no overlap between them was found.

The raw data were processed with the standard STIS pipeline using 
the best reference files. Subsequently, the images were aligned and added up 
with the STSDAS/CRREJ routine which allows for cosmic-rays cleaning. A final 
exposure of 792 s was obtained for the field direct
imaging (Fig.~\ref{f1}). The spectra were taken with the same 50CCD aperture (i.e. no slit was used) 
and the G750L grating whose central wavelength is 7751 \AA. In this set-up, 
the G750L grating is characterised by a spectral resolution of $\sim$ 5 \AA\/ 
per pixel.
The STECF/SLITLESS routine was used to locate the spectra of the brighter 
objects in the field in both the slitless-spectra images. The spectra of three 
galaxies (A, B, and C) and their adjacent backgrounds were 
extracted from each slitless-spectra image with a slit 
of 0.3$''$, 0.8$''$ and 0.5$''$ respectively. Following the background 
subtraction, each set of spectra was manually cleaned from cosmic rays and 
calibrated in wavelength by applying the dispersion correction:
$\lambda$ = (pixel - pixel$_g$)$\times$4.9 + 7751; 
where pixel$_g$ is the X coordinate of each galaxy in the corresponding 
direct image. Only at this point, the spectra of each galaxy were corrected
for the grating response and added up, 
so that the final spectrum of each galaxy has a total exposure time of 2297 s. 

\begin{figure}
\centering
\includegraphics[width=6cm]{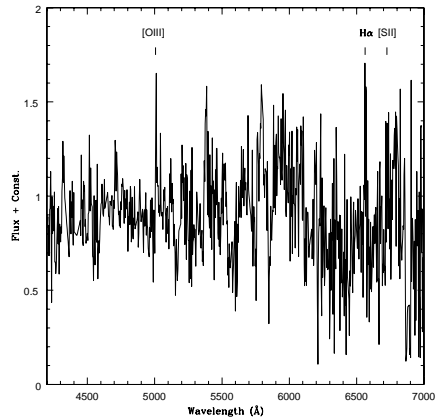}
\caption{Rest frame STIS spectrum of galaxy A in field 204+70.2.}
\label{f2}%
\end{figure}

\begin{figure}
\centering
\includegraphics[width=6cm]{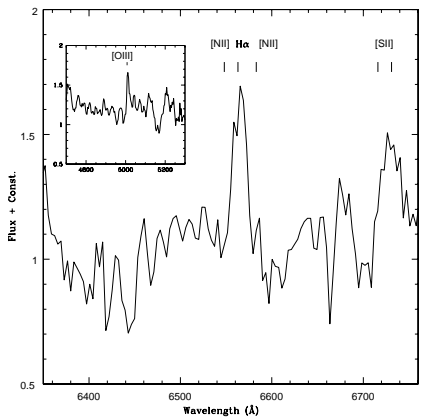}
\caption{Rest frame STIS spectrum of galaxy A in field 204+70.2. Spectrum was smoothed with a 
boxcar of size 3. [OIII]5007, [NII]6548, 6583 and [SII]6716, 6731 are marked. 
H$\alpha$ and [OIII]5007 \AA\ are 3$\sigma$ detections.}
\label{f3}
\end{figure}

The spectra of galaxies B and C have low signal-to-noise to allow any unambiguous
redshift determination. In the spectrum of the brightest object inside box A 
(hereafter, galaxy A) we identified an emission line at 9187\AA\ as H$\alpha$ which corresponds to 
$z=0.4$. Fig.~\ref{f2} and Fig.~\ref{f3} show the 
rest frame spectrum where we have marked the position of the emission lines: 
[OIII]5007, [NII]6548, 6583 and [SII]6716, 6731. However, only H$\alpha$ and
[OIII]5007 \AA\ are 3$\sigma$ detections. 

\section{Discussion}

\begin{figure}
\centering
\includegraphics[width=6cm]{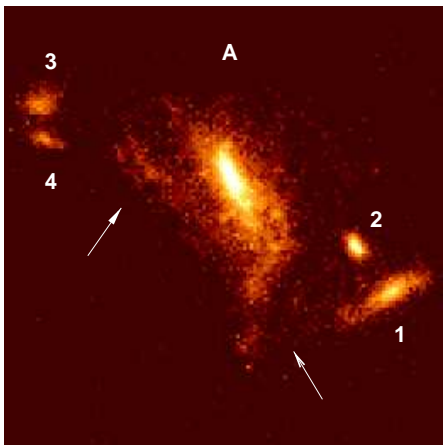}
\caption{Area A of field 204+70.2 (6.5$''$$\times$6.5$''$). Four faint galaxies are identified.
A tail (south) and a disrupted arm (east) are marked with arrows. North is to the top and East is to left.}
\label{f4}
\end{figure}

\begin{figure}
\centering
\vspace{1cm}
\includegraphics[width=5cm]{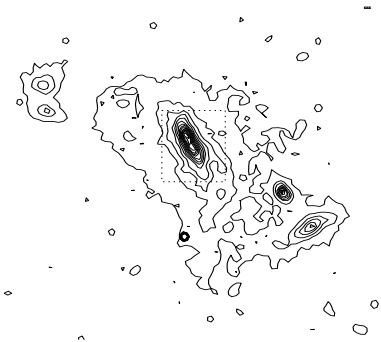}
\vspace{1cm}
\caption{Contour plot of area A of field 204+70.2. The range of the contour levels and the 
number of contours are 53--500 and 12,respectively. STIS background level is 37 $\pm$ 8 counts. 
North is to the top and East is to left.}
\label{f5}
\end{figure}

\begin{figure}
\centering
\vspace{1cm}
\includegraphics[width=6cm]{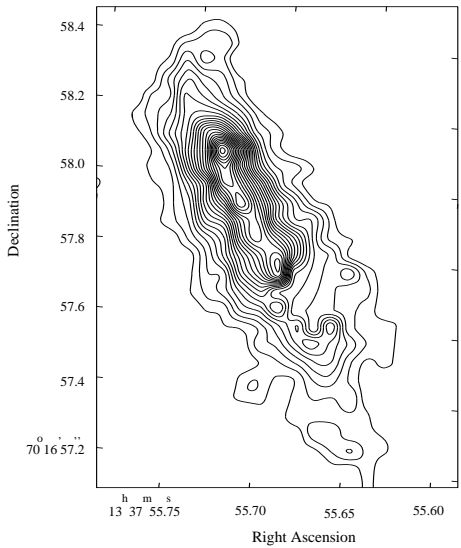}
\vspace{1cm}
\caption{Contour plot of internal region of galaxy A (diameter $\sim$ 5.6 kpc)
- dashed box in Fig.~\ref{f4}. The range of the contour levels and the 
number of contours are  150--900 and 40, respectively. STIS background level is 37 $\pm$ 8 counts. North is to the top and East is to left.}
\label{f6}
\end{figure}

\begin{figure}
\vspace{1.8cm}
\centering
\includegraphics[width=4.5cm,angle=-90]{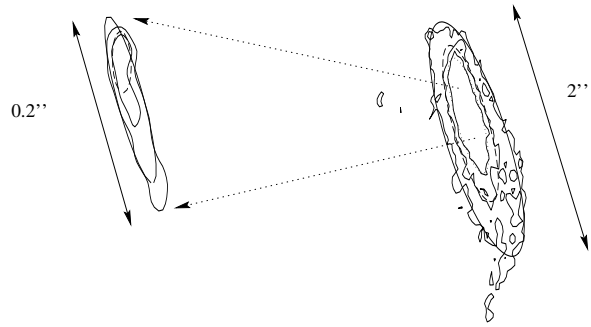}
\caption{Ellipses fitting the contour levels in the inner regions of galaxy A.}
\label{f7}
\end{figure}

The high resolution of STIS allows a detailed view of galaxy A, even though 
it is at a distance of $d_{\rm A}$=885 Mpc ($d_{\rm L}$= 1734 Mpc; 
$H_{\rm o}$=75kms$^{-1}$Mpc$^{-1}$, $q_{\rm o}$=0.5, $\Omega$$_{\rm m}$=1.0).
The presence of a tail on the southern region and a disrupted `arm' on 
the east side are marked by arrows in Fig.~\ref{f4}. The symmetry of the 
morphology of galaxy A was analysed using contour maps (Fig.~\ref{f5} and Fig.~\ref{f6}).
The shape of the contours suggests the presence of a bar along the major axis. We used
the IRAF/STSDAS task ELLIPSE to fit elliptical isophotes to the image (Fig.~\ref{f7}) and obtain 
the structural parameters, ellipticity and position angle 
($\epsilon$ and PA in Table 2). 
The change of 24$^{\rm o}$ in the orientation of the ellipses can be either due to the presence of a 
nuclear bar inclined with respect to the primary bar or to the 
presence of dust absorption and intense sites of star formation (Friedli 1996). 
The latter is probably the case in galaxy A, since two 
opposite hot-spots separated by 0.37$''$ ($\sim$1.6 kpc; 1$''$ = 4.3 kpc at $d_{\rm A}$ = 
885 Mpc) are seeing in the internal region of the galaxy (Fig.~\ref{f6}).

\begin{table}
\caption[]{Structural Parameters of Galaxy A}
\begin{tabular}{ccc}
\hline
Semi-major axis ($''$)  & $\epsilon$ &  PA ($^{\rm o}$)\\
\hline
1.0  & 0.67 $\pm$ 0.02 & 34.15$\pm$ 1.11 \\
0.5& 0.67 $\pm$ 0.01 & 27.72$\pm$ 0.83 \\
0.1& 0.64 $\pm$ 0.07 & 10.09$\pm$ 4.60 \\
\hline
\end{tabular}
\end{table}

To aid in the interpretation of the morphology of galaxy A we used the images generated by 
Hibbard \& Vacca (1997) \footnote{Contour maps were produced using the same procedure
as used for field A. Images were kindly provided by John Hibbard.} in their redshifting 
experiments of peculiar galaxies. We have binned the STIS image in order to have the same 
resolution as WFPC2. The contour maps produced show that the morphology of galaxy A is more 
symmetric and less extended than Arp 299 and NGC 1614. At $z=0.45$ Arp 299 is 9$'' \times 11''$ 
and NGC 1614 is 13$'' \times 11''$ whereas galaxy A is 2.75$'' \times 3.8''$. We have also 
compared galaxy A with NGC 1714 which is a member of Hickson Compact Group 31 and has other physical 
companions in a field of 12$'' \times 9''$. Although the group is significantly larger 
than field A, the contours of the individual galaxies HCG31A (not including HCG31C) 
and HCG31B are similar to galaxy A showing symmetric internal contours and distorted
outer contours.

There are four other small galaxies within field A (number 1 to 4 in Fig.~\ref{f4}); 
however, without  complete redshift information no firm conclusion can be drawn regarding the environment of galaxy A
and they could be chance alignments. If they are at $z=0.4$, galaxy number 1
has diameter $\sim$ 4.3 kpc and the smallest, galaxy number 4, has diameter $\sim$ 2.2 kpc. 
Their sizes are comparable to compact faint blue galaxies which have typical 
$R_{e}$ $<$ 3 kpc (Guzm\'an et al. 1997).

The CLEAR filter covers from 2000 \AA\/ to 1 $\mu$m which limits the astrophysical analysis that
can be done with these data. However, we have estimated the magnitudes of our objects 
by comparing our fields with other images previously observed with HST. We have extracted a 
STIS parallel image which contains a region of the HDF. We have measured the aperture 
magnitudes of the HDF galaxies with the CLEAR filter and compared these values with the cataloged 
HDF total magnitudes (m$_{t}$)
\footnote{The HDF catalog used was taken from the HDF webaddress http://www.stsci.edu/ftp/science/hdf/archive/v2.html}. 
After correction for background, the total counts are divided by the exposure time 
and multiplied by the inverse sensitivity (8.9684505$\times$10$^{-20}$ ergs s$^{-1}$cm$^{-2}$\AA$^{-1}$ counts$^{-1}$).
The zero point of HST (--21.10) was added. CLEAR magnitudes were found to be 
1.053 magnitudes fainter than HDF magnitudes.
CLEAR magnitudes (AB system) for galaxies in region A are given in Table 3.
Galaxy A was recently observed at the Calar Alto 1.23m telescope (F. Comeron
private communication). J, H, and K magnitudes were found to be 18.0 $\pm$ 0.5, 17.8 $\pm$ 0.7, and 
18 $\pm$ 0.4. 

\begin{table}
\caption[]{CLEAR Magnitudes (AB System)}
\begin{tabular}{ccccc}
\hline
Galaxy A & Galaxy 1 & Galaxy 2 & Galaxy 3 & Galaxy 4 \\
\hline
20.84 & 22.90 & 24.47 & 24.71 & 25.15 \\
\hline
\end{tabular}
\end{table}

We have also identified a string of 12 fainter galaxies on the south region of the 
STIS field (box D in Fig.~\ref{f1}). The peak-values of the counts in the faintest galaxies are 
between 3.3 and 37 $\sigma$ detections. The faintest and the brightest galaxies in area 
D have CLEAR magnitudes 26.35 and 23.55, respectively.
The contour maps of 6 of these galaxies 
are typical of interacting spirals with disturbed morphologies. The other 6 galaxies
are too faint to be morphologically classified.

\section{Conclusions}

In this letter we presented the first result of an ongoing project which aim at
compiling a census of the galaxy population at intermediate redshifts.
We are using STIS images taken in parallel mode to select our sample. 
The main advantage of the STIS images is the spatial resolution. Although, the 
exposure times are not long enough to reach the HDF faintest levels in magnitude, the 
high resolution enable us to see intermediate redshift galaxies in more detail. 
We presented a galaxy at $z=0.4$ in which a bar is clearly seen along its major axis. 
The presence of two opposite hot spots separated by $\sim$1.6 kpc are seeing 
in the internal region of the galaxy. The presence of smaller galaxies in the same 
field is also reported. Follow-up observations of this field are planned in order to 
obtain complete redshift information.

\begin{acknowledgements}
We would like to thank N. Fourniol and the HST archive team at ST-ECF for 
their support in retrieving and processing the STIS data. We are grateful to 
our referee, M. Bershady, whose comments and criticisms helped improving our
paper, to F. Schweizer and R. Carlberg for their inspiring comments and 
suggestions, and to J. Hibbard for making their images of peculiar galaxies
available to us. This work was supported by the ESO Director General Discretionary Fund. 
DFM is a `forskarassistent' of Vetenskapsr{\aa}det under project F620-489/2000.
\end{acknowledgements}

\end{document}